\documentstyle[12pt,epsfig]{article}
 1
\def\citenum#1{{\def\@cite##1##2{##1}\cite{#1}}}
\def\citea#1{\@cite{#1}{}}

\def\l{\lambda}

\def\({\left(}
\def\){\right)}

\def\citenum#1{{\def\@cite##1##2{##1}\cite{#1}}}
\def\citea#1{\@cite{#1}{}}

\def\l1vt{\vec{l_{1\perp}}}

\def\bt{b_{\perp}}

\def\bt2{$b^2_t$}

\def\jol1{$J_0(\,l_{1\perp}\,r_{\perp}\,)$}

\def\citea#1{\@cite{#1}{}}

\def\beq{\begin{equation}}
\def\eeq{\end{equation}}
\def\bea{\begin{eqnarray}}
\def\eea{\end{eqnarray}}

\def\bbbz{{\mathchoice {\hbox{$\sf\textstyle Z\kern-0.4em Z$}}
{\hbox{$\sf\textstyle Z\kern-0.4em Z$}}
{\hbox{$\sf\scriptstyle Z\kern-0.3em Z$}}
{\hbox{$\sf\scriptscriptstyle Z\kern-0.2em Z$}}}}
\relax
\begin{document}
\newcounter{savefig}
\newcommand{\alphfig}{\addtocounter{figure}{1}%
\setcounter{savefig}{\value{figure}}%
\setcounter{figure}{0}%
\renewcommand{\thefigure}{\mbox{\arabic{savefig}-\alph{figure}}}}
\newcommand{\resetfig}{\setcounter{figure}{\value{savefig}}%
\renewcommand{\thefigure}{\arabic{figure}}}

\begin{titlepage}
\noindent
\begin{flushright}
March   1998 \\ 
TAUP 2485-98\\
{\tt hep-ph/}
\end{flushright}
\vspace{1cm}
\begin{center}
{\Large \bf {   ENERGY DEPENDENCE OF THE SURVIVAL PROBABILITY OF  
  LARGE RAPIDITY GAPS }\\[4ex]}

{\large E. G O T S M A N${}^{a), 1)}$, E. L E V I N${}^{a),b),2)}$\,\,
 and U.\,\,M A O R${}^{a), 3)}$}
 \footnotetext{$^{1)}$ Email: gotsman@post.tau.ac.il .}
\footnotetext{$^{2)}$ Email: leving@post.tau.ac.il.}
\footnotetext{$^{3)}$ Email: maor@post.tau.ac.il.}\\[4.5ex]
{\it a) School of Physics and Astronomy}\\
{\it Raymond and Beverly Sackler Faculty of Exact Science}\\
{\it Tel Aviv University, Tel Aviv, 69978, ISRAEL}\\[1.5ex]
{\it b)  Theory Department, Petersburg Nuclear Physics Institute}\\
{\it 188350, Gatchina, St. Petersburg, RUSSIA}\\[3.5ex]
\end{center}
~\,\,\,
\vspace{2cm}

{\large \bf Abstract:}

\par The energy dependence for the survival probability of large rapidity
gaps (LRG) \\
 $< \mid S \mid^2 >$,  is calculated in an Eikonal model assuming a
Gaussian opacity. The parameters determining  
$< \mid S \mid^2 >$   are evaluated directly from
experimental data, without further 
recourse to models. We find that  $< \mid S \mid^2 >$ decreases with
increasing energy, in line with recent results for LRG dijet production at
the Tevatron.

\end{titlepage}

\subsection*{ \bf General approach}
\par Large rapidity gaps (LRG) have been studied both at the Tevatron
\cite{CDF}\cite{D0},and at HERA \cite{DERR}.
LRG  are expected whenever we have a process where a colour singlet
 is exchanged in the t channel. A prolific amount of data on LRG
has been accumulated for dijet production, which is the channel that we
will concentrate on in this letter.

\par Historically, both Bjorken \cite{Bj} and Dokshitzer et al. \cite
{Dok}, suggested utilizing rapidity gaps as a signature for Higgs
production, in the W-W fusion process in hadron-hadron collisions. As
pointed out \cite{Bj}, the fraction of events for which a LRG is
expected, is the product of two factors (see Fig.1)
\beq \label{1}
f_{gap} = < \mid S \mid^2 > \cdot F_{s}
\eeq
\begin{figure}[defgap]
\epsfig{file=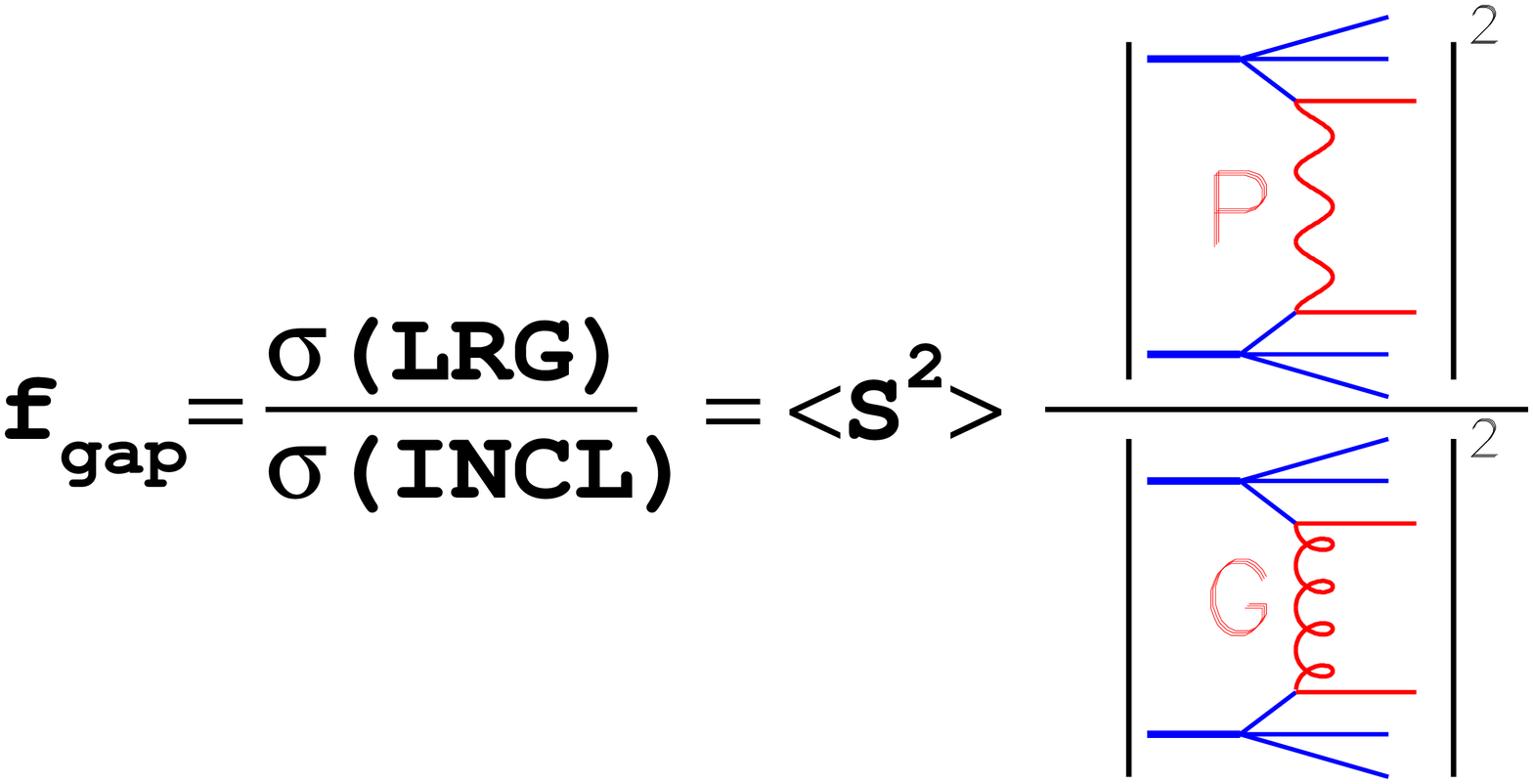,width=140mm}
\caption{ Pictorial definition of $f_{gap}$, where P and G represent
respectively, the exchange of a colour singlet and a colour octet }
\label{Fig.1}
\end{figure}
where $< \mid S \mid^2>$ denotes the survival probability, i.e. the
fraction of events for which the spectator interactions do not fill the
rapidity gap of interest.  $F_{s}$  is the  fraction of events due to
t-channel singlet exchange.
Bjorken \cite{Bj} by comparing the rate for the exchange of two gluons, in
a
colour singlet state, to the exchange of one gluon in a colour octet
state, estimated
$F_{s}$ to be about 15 \%. Other authors \cite{GLM1} \cite{BH} \cite{EGH}
 \cite{Del} have suggested alternative ways to estimate $F_{s}$, with
results
that  are not too different from that of \cite{Bj}.  

\par In this note we only concern ourselves  with evaluating the first
term in eq.(1) i.e. the survival probability. Experimentalists use a more
practical definition of $f_{gap}$. At the Tevatron $f_{gap}$ for 
LRG is defined to
be the ratio:
$$f_{gap}\; = \;  \frac{cross\;\; section\;\; for\;\; dijet\;\;
production \;\; with \;\;
LRG}{inclusive \;\; cross \;\;
section \;\;  for\;\;  dijet \;\; production} $$
The jets are required to have a transverse energy $E_{T}^{jet} >\; 12\; 
GeV$
at DO, and $ E_{T}^{jet} >\; 20\; GeV $ at CDF.

\par   The survival probability is easily
defined using the eikonal model in impact parameter space. Normalizing
our amplitude so that
$$ \frac{d \sigma}{dt} =\pi\; \mid f(s,t) \mid^{2}$$    
\beq \label{2}
\sigma_{tot} = 4 \pi Im f(s,0)   
\eeq
The amplitude in impact parameter space is then given by:
\beq \label{3}
a_{el}(s,b) = \frac{1}{2 \pi} \int d^{2}q e^{-i \vec{q} \cdot \vec{b} }
f(s,t)
\eeq
from which we have

$$\sigma_{tot} = 2 \int d^ {2}b Im a_{el}(s,b) $$
\beq \label{4}
\sigma_{el} =  \int d^{2}b \mid a_{el}(s,b) \mid^{2}
\eeq
s-channel unitarity implies that $\mid a(s,b) \mid \  \leq$ 1, and when
written in diagonalized form we have
\beq \label{5}
2 Im a_{el}(s,b) = \mid a_{el}(s,b) \mid^{2} + \; G_{in}(s,b)
\eeq
where the inelastic cross section
\beq \label{6}
\sigma_{in} = \int  d^{2}b G_{in}(s,b)
\eeq
s channel unitarity is most easily enforced in the eikonal approach where,
assuming that $a(s,b)$ is purely imaginary, we have
$$a_{el}= i[1 - e^{- \Omega(s,b)/2}]$$
\beq \label{7}
G_{in}(s,b) = 1- e^{- \Omega(s,b)}
\eeq
where  $ \Omega(s,b)$ is a real function.

Hence $P(s,b) = e^{- \Omega(s,b)}$ is the probability that no inelastic
interaction takes place at impact parameter b.

\par It is expedient to assume that the opacity $ \Omega(s,b)$ can be 
approximated by a Gaussian
$$ \Omega(s,b) =\nu(s)\Gamma(s,b)$$
where 
\beq \label{8}
  \nu(s) \;\; = \;\; \Omega(s,b = 0) 
\eeq
with the profile
\beq \label{9} 
 \Gamma(s,b) =\frac{1}{\pi R^{2}(s)}e^{-\frac{b^{2}}{R^{2}(s)}}
 \eeq
The Gaussian input eq.(9) corresponds to an exponential behaviour in
t-space \cite{GLM2}. 
In the Regge framework the "soft" radius is of the form
\beq \label{10}
R^{2}_{S}(s) = 4\; R^{2}_{0} + 4\; \alpha^{'}(0) ln(\frac{s}{s_{0}}) \; =
2\; 
B_{el}(s)
\eeq
where $B_{el}$(s) denotes the forward slope of  the elastic cross section
(i.e. 
$ \frac{d \sigma}{dt} \sim e^{B_{el}(s)t}$).

\par With the amplitude in  this form, the integrations in eq.(4)
can be done analytically, and we obtain the following closed
expressions for
\beq \label{11}
\sigma_{tot} = 2 \int d^{2}b (1 - e^{- \Omega(s,b)/2}) 
= 2 \pi R^{2}(s)[ln(\frac{\nu(s)}{2}) + C - Ei(- \frac{\nu(s)}{2})]
\eeq
\beq \label{12}
 \sigma_{in} = 2 \int d^{2}b (1 - e^{- \Omega(s,b)})
=  \pi R^{2}(s)[ln(\nu(s)) + C - Ei(- \nu(s))]   
\eeq
\beq \label{13}
 \sigma_{el} = \sigma_{tot} - \sigma_{in}  
=  \pi R^{2}(s)[ln(\frac{\nu(s)}{4}) + C + Ei(- \nu(s)) - 2 Ei(-
\frac{\nu(s)}{2)} ]     
\eeq
where
$$ Ei(x) = \int_{- \infty}^{x} \frac{e^{t}}{t} dt $$
and C = 0.5773.

\par Note that the ratio $ \frac{\sigma_{el}}{\sigma_{tot}} $ only depends 
on the factor $ \nu(s)$.

\par Following \cite{Bj}  the survival probability (see eq.(1))
$< \mid S \mid^2 >$ is defined
as the normalized product of two quantities.
The first is the convolution over the parton densities, of two interacting
projectiles presenting the cross section for the  hard parton-parton
collisions (dijet production) under discussion. The second term is P(s,b)
i.e. the probability that no inelastic interaction takes place in the
rapidity interval of interest. 

\par In the eikonal formalism 
\beq \label{14}
< \mid S \mid^2> = \frac{ \int d^{2}b \Gamma_{H}(b) P(s,b)}
{ \int d^{2}b \Gamma_{H}(b)} 
\eeq

where $P(s,b) = e^{- \Omega(s,b)}$, and the "hard" profile
\beq \label{15}
\Gamma_{H}(b) = \frac{1}{\pi R^{2}_{H}(s)}e^{-\frac{b^{2}}{R^{2}_{H}(s)}}
 \eeq
where $R_{H}$ denotes the radius of interactions in the "hard" scattering
process.

\par We define
\beq \label{16}
a(s) =\frac{R^{2}_{S}(s)}{R^{2}_{H}(s)}
\eeq
i.e.
$$ a(s) = \frac{\pi R^{2}_{S}(s)}{ \pi R^{2}_{H}(s)} = \frac{
interaction\;\;
area\;\; for \;\; soft \;\; collisions}{interaction\;\; area\;\; for \;\;
hard \;\; collisions} > \;\;1 $$
Evaluating eq.(14) yields
\beq \label{17}
< \mid S \mid^2> =  \frac{a(s) \gamma[a(s),
\nu(s)]}{[\nu(s)]^{a(s)}}
\eeq
where the incomplete gamma function
$ \gamma(a,x) = \int_{0}^{x} z^{a-1}e^{-z}dz $

\subsection*{ \bf Numerical evaluation of $ < \mid S \mid^2> $}

\par Previous estimations  of the LRG survival probability \cite{Bj}
\cite{GLM1} were derived from  various models assumed to
describe the "soft" and "hard"  interactions. Our approach here is to
determine the
parameters on which $ < \mid S \mid^2> $ depends  in the Eikonal
Model, directly from the
relevant experimental data  without further recourse to models.

\par The survival probability is  sensitive to the value taken for
 the radius of the ``hard" interaction $R_H$.
 We wish to evaluate this radius directly from the
experimental data, namely, from the cross section of the double parton
interaction measured by CDF at the Tevatron \cite{CDFDP}. Indeed, the CDF
collaboration  have measured the process of inclusive production  of two
``hard" jets with almost compensating  transverse momenta in each pair, 
and
with  values of rapidity that are very similar.
Such pairs can only be produced   in   double parton collisions, and
their
cross section can be calculated using the Mueller diagram given in Fig.2.

The  double parton scattering cross section can be written in
the form ( see Fig.2) \cite{CDFDP}:

\beq \label{DP}
\sigma_{DP}\,\,=\,\,m\,\frac{\sigma_{incl}(2jets)\,
\sigma_{incl}(2jets)}{2\,\sigma_{eff}}\,\,,
\eeq
where the factor $m$ is equal to  2 for different pairs of jets, and to
1  for identical pairs.  The experimental value of
$\sigma_{eff}\,=\,14.5\,\pm\,1.7\,\pm\, 2.3 \, $ mb. In  Fig.2,
$\sigma_{eff}$ is obtained by integrating over momentum transfer
$\vec{q}$,  or over the impact parameter $b$, namely
\beq \label{DP1}
\frac{1}{\sigma_{eff}}\,\,=\,\,\int\,d^2 b \,\Gamma_H^2(b)\,\,=\,\,
\frac{1}{2 \,\pi\,R^2_H}\,\,.
\eeq
Comparing with the experimental value of $ \sigma_{eff}$, one obtains
$R^2_H\,\,= 5 - 7 \,GeV^{-2}$.

\par Recent data from HERA \cite{AL} indicate that the photo and electro
production of J/$\psi$ is a "hard" (short distance) process. Data
 has $ \sigma_{H}(\gamma + p \rightarrow  J/ \psi + p ) \sim W^{0.8} $
while the analagous "soft" process
$ \sigma_{s}(\gamma + p \rightarrow  \rho + p ) \sim W^{0.22}$  .
The forward slope
 $B_{el}(\gamma^{*} + p \rightarrow  J/ \psi + p ) \sim 4 \; GeV^{-2} $,
while for the inelastic diffraction process
$\gamma^{*} + p \rightarrow  J/ \psi + X $,
$B_{in}(\gamma^{*} + p \rightarrow  J/ \psi + X ) \sim 1.6 \; GeV^{-2}
$, both $B_{el}(s)$ and $B_{in}(s)$
 appear to be independent of $Q^{2}$ and W, as one would expect for a
"hard" process. This experimental observation may suggest that the proton
is a two radii object \cite{GLM3}. For the present calculation we use an
average radius which we estimate to be
$R^{2}_{H} \;=\;  8 GeV^{-2} $ . This value  is  smaller than than has
been
assumed up to now \cite{Bj} \cite{GLM1}.
This is a conservative choice, and it is possible that future experimental
data will  dictate an even smaller value
 of $R^2_H \,=\,5 - 7\; GeV^{-2}$,  as 
 the CDF data for the double parton cross section suggests.

 \begin{figure}[exnu]
\epsfig{file=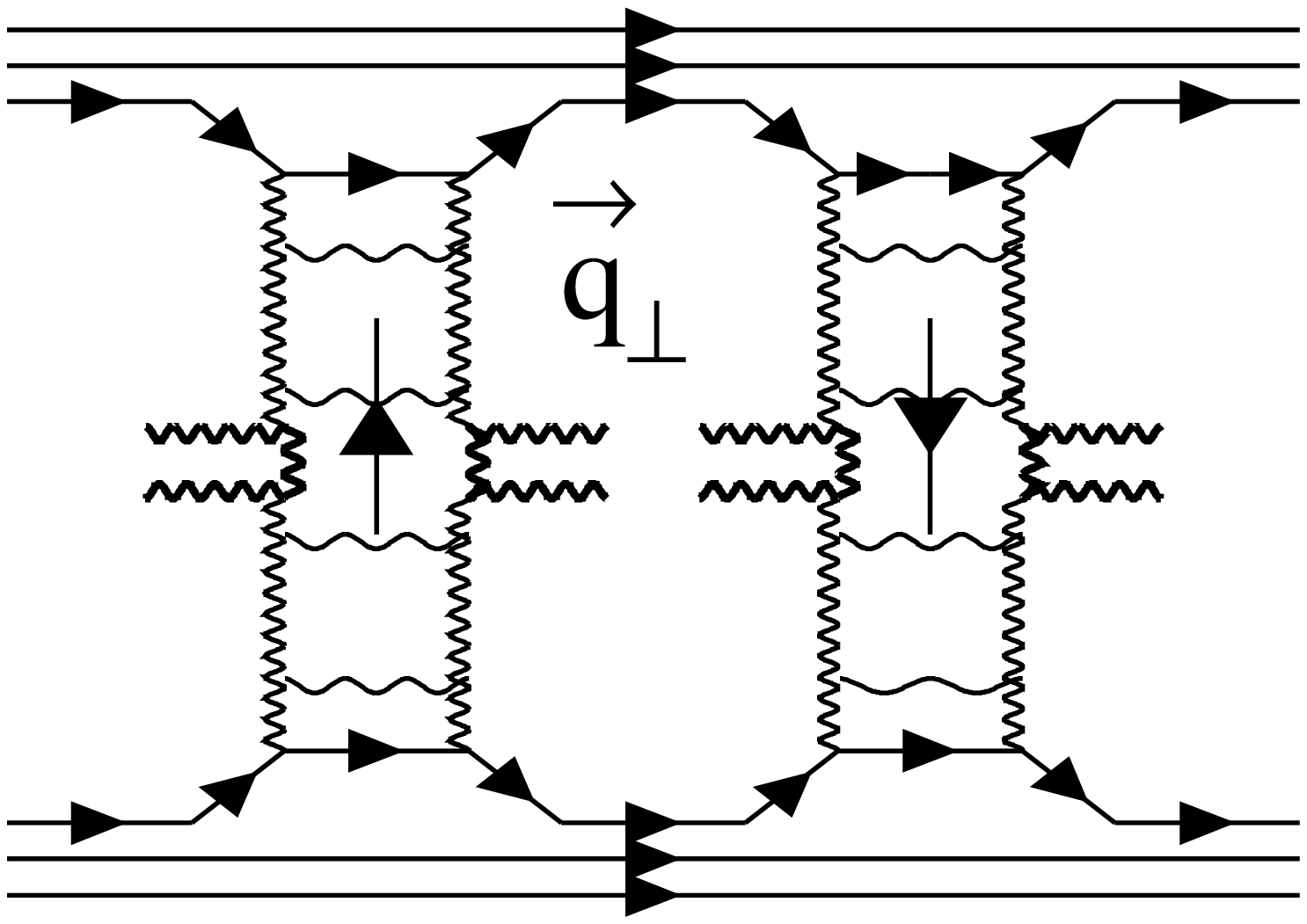,width=140mm}
\caption{ The Mueller diagram for the double parton interaction. .}
\label{Fig.2}
\end{figure}

\par The pp and $\bar{p}$p channels have measurements of 
$B_{el}(s) \; = \; ( \frac{1}{2}\;  R^{2}_{s} )$ and 
$\frac{\sigma_{el}}{\sigma_{tot}}$ over a large energy range \cite{DUR}.
 The parameter 
$ a(s) = \frac{R^{2}_{S}(s)}{R^{2}_{H}}$ can thus
be determined directly from experimental data.
 The second parameter which is required to evaluate $ < \mid S \mid^2>
$ is $ \nu(s)$ (see eq.(8)). Using experimental values
from pp and $\bar{p}$p scattering data \cite{DUR} for the
ratio $ \frac{\sigma_{el}}{\sigma_{tot}} $ we can find the corresponding
value of $ \nu(s)$ at the appropriate value of s ( energy squared),
using eqns.(11) and (13).
The dependence of $ \nu(s)$ as a function of the ratio
 $ \frac{\sigma_{el}}{\sigma_{tot}} $ for $ p \bar{p}$  scattering is
shown in fig.3.

\begin{figure}[exnu]
\epsfig{file=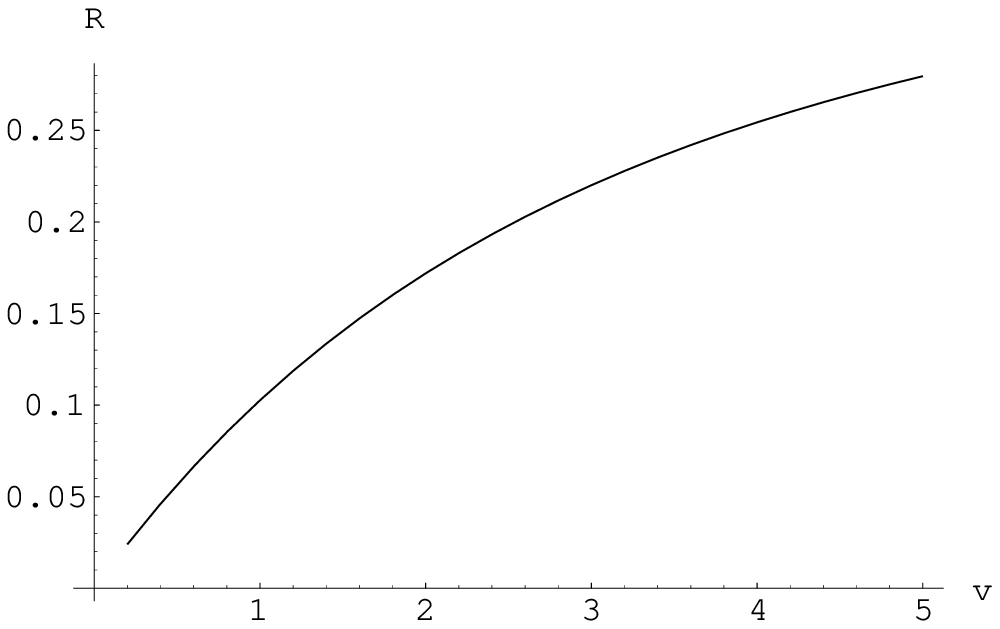,width=140mm}
\caption{ $ R^{'} =   \frac{\sigma_{el}}{\sigma_{tot}} $ 
 as a function of $ \nu(s)$ .}  
\label{Fig.3}
\end{figure}

\par Once we have determined $\nu(s)$ and a(s) (eq.(16))
, the survival probability can be determined from  
eq.(17). In Fig.4 we display  $ < \mid S \mid^2> $ as a function of
 $\nu(s)$ and a(s).
In Fig.5 we show a contour plot for  $ < \mid S \mid^2> $, for
various values of $\nu(s)$ and a(s).

\begin{figure}[snua]
\epsfig{file=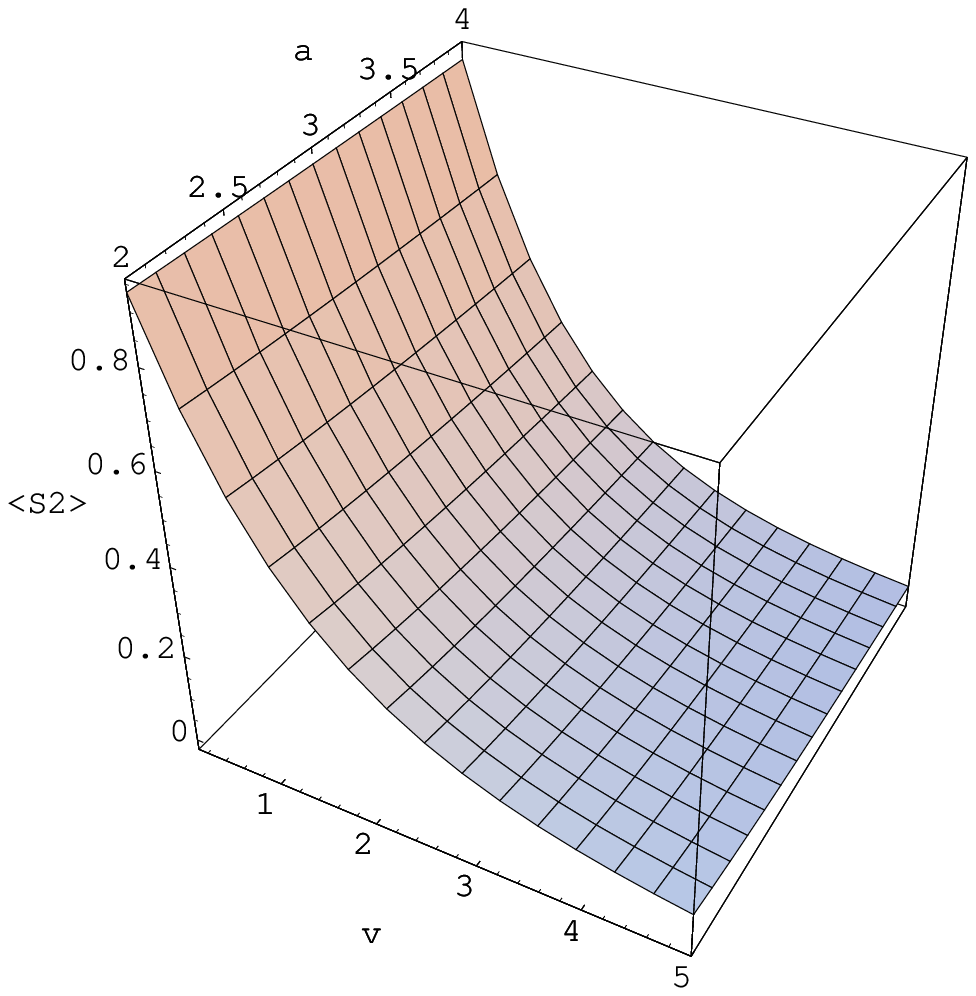,width=140mm,height= 65mm}
\caption{ Survival probability as a function of $\nu(s)$ and a(s)}.
\label{Fig.4}
\end{figure}

\begin{figure}[cont]
\epsfig{file=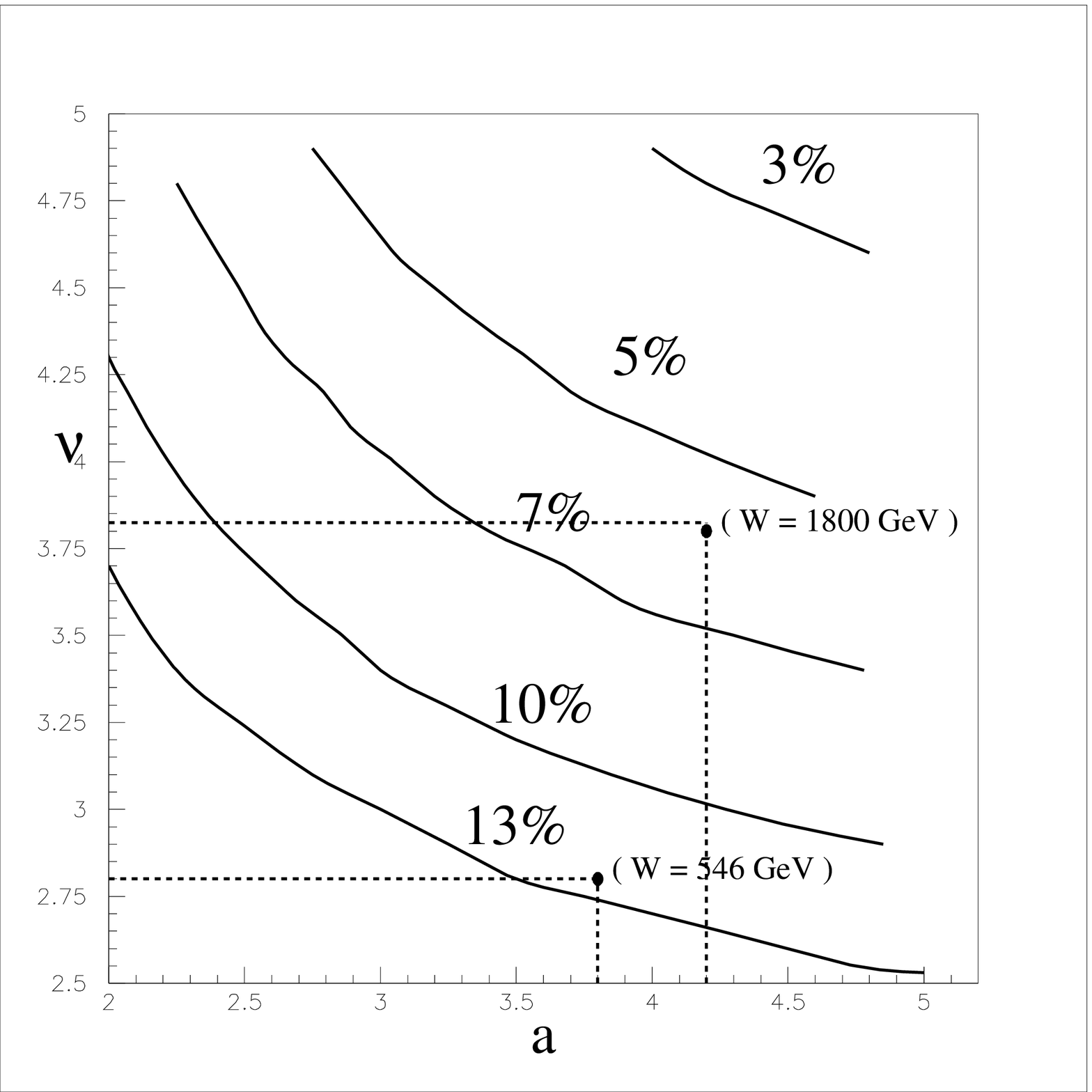,width=140mm}
\caption{ Contour plot of survival probality $ < \mid S \mid^2> $.}
\label{Fig.5}
 \end{figure}

\subsection*{ \bf Comparison with data}
\par Recently D0 \cite{Brandt} have measured LRG in dijet production
in $\bar{p}$p scattering at center of mass energies of $\sqrt{s}$ = 630
and 1800 GeV, and find
$ f_{gap}(\sqrt{s}\; =\; 630 GeV) \; = \; 1.6 \pm \; 0.2\%$
and $f_{gap}(\sqrt{s}\; =\; 1800 GeV) \; = \; 0.6 \pm \; 0.1\%$
\footnote{Prelimenary results presented by J. Perkins \cite{Perkins}
at the LISHEP98 Workshop reported a new value of R = 3.4 $\pm\;$ 1.3}
i.e.
$$  R = \frac{f_{gap}(\sqrt{s} = 630\; GeV)}
{f_{gap}(\sqrt{s}\;=\; 1800\; GeV)} \;\;= \;\; 2.6 \;\;\pm 0.6$$
 The ZEUS collaboration \cite{DERR} reported a value of
$f_{gap}  \;\; = 7\;\pm 2\;\%$, at a center of mass energy $<W> \sim$ 150
GeV in the photoproduction of dijets.

\par The survival probability  $ < \mid S \mid^2> $ increases as we go to
lower energies, which corresponds to lower values of $\nu(s)$ and a(s)
(see fig.4). In fig.5 we have marked the values of $\nu(s)$ and a(s) which
correspond to $\sqrt{s}$ = 546 and 1800 GeV. We would like to stress
that the input data used in calculating  $ < \mid S \mid^2> $ was taken
from p-p and $\bar{p}$-p interactions. Our predictions for the LRG
survival probability are: 
\begin{tabbing}
xxxxxxxxxxxxxxxxxxxxx     \=\kill
 
  $\sqrt{s}$\,\, (GeV)    \> $ < \mid S \mid^2> $ \\ 
   150         \>   16.3  \,\%\\
   630       \>   12.8 \,$ \pm$ \,0.4 \% \\
   1800       \>    5.9 \,$ \pm $\, 0.4 \%\\
\end{tabbing}
Hence the ratio of: 
$$ \frac{< \mid S \mid^2>_{\sqrt{s}=630}}
{< \mid S\mid^2>_{\sqrt{s}=1800}} \;\;= \;\; 2.2\;\pm 0.2 $$

\section*{ \bf Conclusions}
Recent measurements at HERA \cite{DERR} provide us, for the first time,
with a  value for the radius of the "hard" interactions i.e. about 
8  GeV$^{-2}$. Assuming a one radius model for the proton, and a Gaussian
form for the profile in impact parmater space, gives us an explicit
expression for  $ < \mid S \mid^2> $.
\par  We note  that the LRG survival probability  $ < \mid S \mid^2> $
decreases with increasing energy (see Fig.5). This is a consequence of the
fact that experimental data has both $\frac{\sigma_{el}(s)}{\sigma_{tot}}$
and $B_{el}(s)$ increasing with s.
This trend is consistent with the observations of the D0 collaboration
\cite{Brandt}\cite{Perkins} who find
$$\frac{f_{gap}(\sqrt{s}=630 GeV)}{f_{gap}(\sqrt{s}=1800 GeV)}
\; = \; 3.4 \pm \; 1.3 $$
 Using exprimental data to 
directly fix the parameters $\nu(s)$ and a(s) we find
$$ \frac{ < \mid S \mid^2>_{\sqrt{s} = 546 GeV}}
{ < \mid S \mid^2>_{\sqrt{s} = 1800 GeV}} \; = 2.3 \; \pm 0.2 $$ 
 which we can interpolate to values measured at  Tevatron energies
i.e.
$$ \frac{ < \mid S \mid^2>_{\sqrt{s} = 630 GeV}} 
{ < \mid S \mid^2>_{\sqrt{s} = 1800 GeV}} \; = 2.2 \; \pm 0.2 $$

This is consistent with  $F_{s}$, the singlet to octet colour exchange
contribution
(see eq.(1)) being  mildly energy dependent as suggested by
Eboli et al. \cite{EGH},
or energy independent as proposed by   Buchm\"{u}ller and  Hebecker 
\cite{BH}. 
 
\par  The main results of this paper, that the survival
probability decreases with increasing energy, and  the   predicted rates, 
are   in agreement with what is  observed at the Tevatron.

{\bf Acknowlegements:}
This research was supported in part by the Israel Science Foundation
founded by the Israel Academy of Science and Humanities.


\begin{thebibliography}{99}
\bibitem{CDF}
CDF  Collaboration; F. Abe et al., Phys. Rev. Lett. {\bf 74}  (1995) 855.
\bibitem{D0}
D0 Collaboration,  S. Abachi et al., Phys. Rev. Lett. {\bf 72} 
(1994) 2332; Phys. Rev. Lett. {\bf 76} (1996) 734.
\bibitem{DERR}
ZEUS collaboration, M. Derrick et al., Phys. Lett. {\bf B315} (1993) 481;
Z. Phys. {\bf C68} (1995) 569; Phys. Lett. {\bf B369} (1996) 55.
 \bibitem{Bj}
 J. D. Bjorken, Int. J. Mod. Phys.{\bf A7} (1992) 4189; Phys. Rev. {\bf
D47} (1993) 101.
\bibitem{Dok}
 Yu. L. Dokshitzer, V. Khoze and T. Sjostrand,
 Phys. Lett. {\bf B274}, (1992) 116. 
\bibitem{GLM1}
E. Gotsman, E.M. Levin and U. Maor, Phys. Lett. {\bf B309}, (1993) 199.
\bibitem{EGH}
O. J. P. Eboli, E. M. Gregores and F. Halzen, preprint MADPH-97-995
(1997), [hep-ph/9708283].
\bibitem{BH}
W. Buchm\"{u}ller and A. Hebecker, Phys. Lett. {\bf B355} (1995) 573.
\bibitem{Del}
V. Del Duca and W. K. Tang, Phys. Lett. {\bf B312} (1993) 225.
\bibitem{GLM2}
E. Gotsman, E.M. Levin and U. Maor, Z. Phys.  {\bf C57}, (1993) 677.
\bibitem{CDFDP}
CDF Collaboration, F.Abe et al.: FERMILAB-Pub-97/083-E.
\bibitem{AL}
A. Levy,  preprint TAUP 2468-97      [hep-ph/9712519  ].
\bibitem{GLM3}
E. Gotsman, E.M. Levin and U. Maor, Phys. Lett.{\bf B403} (1997) 120.
\bibitem{DUR} 
 HEPDATA, Durham/RAL HEP Databases.
\bibitem{Brandt} 
A. Brandt at  Workshop on "Interplay between Soft and Hard
interactions in DIS", Heidelberg Germany, September 1997.
\bibitem{Perkins}
J. Perkins at "LAFEX Int. School on High Energy Physics", Rio de Janerio,
February 1998.  
\end{thebibliography}
\end{document}